# Harnessing Retrieval-Augmented Generation (RAG) for Uncovering Knowledge Gaps

Joan Figuerola Hurtado

Independent Researcher

joanfihu@gmail.com

## Abstract

We present a methodology for uncovering knowledge gaps on the internet using the Retrieval Augmented Generation (RAG) model. By simulating user search behaviour, the RAG system identifies and addresses gaps in information retrieval systems. The study demonstrates the effectiveness of the RAG system in generating relevant suggestions with a consistent accuracy of 93%. The methodology can be applied in various fields such as scientific discovery, educational enhancement, research development, market analysis, search engine optimization, and content development. The results highlight the value of identifying and understanding knowledge gaps to guide future endeavours.

## 1 Introduction

The increasing number of users dissatisfied with the relevance of commercial search engine results is surprising, given the unprecedented access to vast information and sophisticated search technologies [1, 2].

In this paper, we employ the Retrieval Augmented Generation (RAG) model to simulate user search behaviour, aiming to identify and address knowledge gaps on the Internet. We posit that uncovering and bridging these gaps is crucial for enhancing the efficacy of information retrieval systems.

## 2 Related Work

Yom. et. al [14] presents an algorithm to estimate query difficulty. Estimation is based on the agreement between the top results of the full query and the top results of its sub-queries. In doing so, difficult queries reveal gaps in a content library. The methodology is based on training an estimator based on a small dataset. We argue that there are now simpler LLM prompting techniques that do not require training a custom model and yield better generalisation across multiple domains.

A Large Language Model (LLM) [4] generates text-based responses, while RAG [3] is an AI framework used to enhance the quality of LLM-generated responses by grounding them on external sources of knowledge. These technologies combine to provide accurate, up-to-date information and improve the generative process of language models.

## 3 Methodology

To identify knowledge gaps, we simulate user interactions with search engines in a structured process. Initially, we begin with a query and methodically review each search result until an answer is found. If the first top 10 results do not yield an answer, we generate up to four alternative queries and retrieve up to two documents per query, iterating through the search process again.

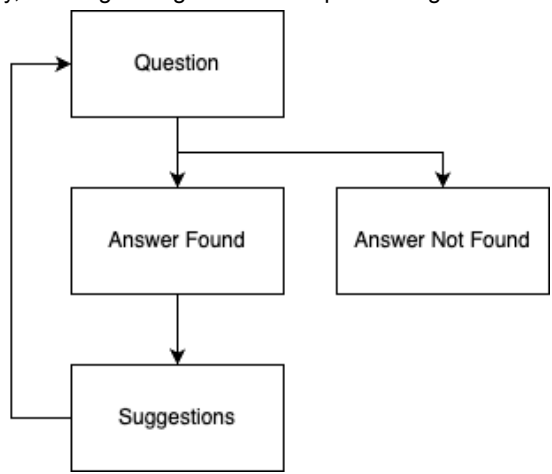

Figure 1: **Iteration loop to find knowledge gaps**

Our approach utilises AskPandi [12], a Retrieval-Augmented Generation (RAG) system, to mimic user behaviour. AskPandi integrates Bing's web index for data retrieval and GPT as a reasoning engine. After finding an answer, we capitalise on the in-context capabilities [5, 6, 7] of LLMs to generate a series of relevant follow-up questions. This process is guided by the premise that a well-generalised [8] LLM should provide useful recommendations based on the initial question and answer. The prompt we use is:

> 'Based on the answer '{}' and the question '{}', what are some potential short follow-up questions?'

This methodology diverges from traditional recommender systems [9], which filter through existing content. In contrast, our system focuses on generating the most relevant content, regardless of its preexistence, highlighting a shift from extractive to generative approaches. The process is then iterated, with each cycle going deeper into the query's topic, thus increasing the difficulty of finding relevant information. We consider the emergence of a knowledge gap when the LLM can no longer generate an answer.

In terms of terminating the process, we incorporate a mechanism to identify stop words in answers. We explored two methods: either letting the model naturally produce a stop word or directing the model to generate one in cases of uncertainty [10].

This comprehensive process not only helps in identifying knowledge gaps but also enhances our understanding of the potential of generative AI in facilitating more relevant information retrieval systems.

## 4  Experiments

We build a dataset with 500 search queries classified in 25 categories. We pick the parent categories from Google Trends as of 2023 [11]. Given that Google Trends derives its data from Google search queries, it is hypothesised that this tool provides a representative sample of the general online search behaviour. All the 500 search queries can be found in our GitHub repository [13].

1. Arts & Entertainment
2. Autos & Vehicles
3. Beauty & Fitness
4. Books & Literature
5. Business & Industrial
6. Computers & Electronics
7. Finance
8. Food & Drinks
9. Games
10. Health
11. Hobbies & Leisure
12. Home & Garden
13. Internet & Telecom
14. Jobs & Education
15. Law & Government
16. News
17. Online Communities
18. People & Society
19. Pets & Animals
20. Property
21. Reference
22. Science
23. Shopping
24. Sports
25. Travel

For each category, we generate 20 queries grouped by their complexity: easy and difficult. To determine the complexity of each query, we use the following methodology:

Length of Query
- Easy: Short queries, usually 1-3 words.
- Difficult: Very long queries or full sentences, more than 6 words.

Specificity of Query
- Easy: General or broad queries.
- Difficult: Highly specific, niche, or detailed queries.

Use of Jargon or Technical Terms
- Easy: Common language, no specialised terms.
- Difficult: Heavy use of technical terms, jargon, or acronyms.

Ambiguity or Clarity of Query
- Easy: Clear and straightforward, with likely one main interpretation.
- Difficult: Ambiguous, requiring context or additional information to interpret.

Search Intent
- Easy: General information seeking or popular topics.
- Difficult: In-depth research, controversial topics, or highly detailed queries.

Knowledge Level Required
- Easy: Suitable for a general audience, no special knowledge needed.
- Difficult: Requires in-depth knowledge or expertise in the field.

Query Format
- Easy: Basic questions or keyword searches.
- Difficult: Complex questions, hypotheticals, or requiring multi-step thinking.

For each search simulation, we measured the following metrics:
- Accuracy: the percentage of queries that were answered correctly by the RAG system. Answers that have been manually reviewed.
- Topic Depth: the number of iterations until the LLM system stopped answering the question.

- Average number of sources used per search simulation.

## 5 Analysis

We carried out search simulations for 60 keywords, generating 323 answers across 655 sources. We have found that using more than 60 keywords from the initial 500 keywords dataset did not make a significant difference. All the search simulations can be found in our GitHub repository [13]. The results demonstrate the effectiveness of using a RAG system in simulating user search behaviour and generating relevant suggestions.

With a consistent accuracy of 93% for both simple and complex keywords, the RAG system proved to be a reliable tool for information retrieval. The study also found that finding sources becomes slightly more challenging for specific topics, as indicated by the average number of sources needed per keyword difficulty, 10.9 sources for easy queries and 11.23 for difficult ones. No significant differences were observed in accuracy or source quantity across categories, likely due to the broad and balanced nature of the selected categories.

Additionally, we discovered that on average, a knowledge gap is encountered at the fifth level of topic depth. This suggests that the internet may have limitations in providing in-depth information on certain subjects. Our methodology effectively highlights these knowledge gaps, showing a straightforward approach to identifying them in various topics.

## 6 Applications

Recommending nonexistent content is a powerful tool for revealing knowledge gaps. This approach has a wide range of applications, including:

1. Scientific Discovery: It can pinpoint unexplored areas in research, highlighting future research topics that have yet to be investigated.
2. Educational Enhancement: By identifying missing elements in learning materials, it helps in creating more comprehensive educational resources.
3. Research Development: This method can uncover untapped research opportunities, guiding scholars and scientists towards novel inquiries.
4. Market Analysis: In the business realm, it can reveal product gaps in a catalogue, offering insights for new product development.
5. Search Engine Optimization: Improving search recommendations by identifying what users might be looking for but isn't currently available online.
6. Content Development: It aids in recognizing content gaps within a content library, assisting content creators in filling these voids.

Each of these applications demonstrates the value of identifying and understanding what is missing, thereby guiding future endeavours in various fields.

## 7 Conclusion

We have successfully demonstrated a methodology for identifying knowledge gaps in content libraries. For future work, there is potential to expand this research by exploring alternative search simulation methods. Specifically, utilising agents could be a promising avenue. These agents, with their broader bandwidth in search engine usage and content processing, offer capabilities surpassing those of human users. Future research could extend the evaluation to additional answer engines, thereby enabling a more comprehensive benchmarking of the estimation methodology outlined in reference [14].

It's worth pointing out that we don't have direct access to a web index to do a more rigorous evaluation. Future work could consider the system's ability to predict whether a query is a MCQ (missing content query) [14] given gold-standard labels (perhaps using a TREC-style test collection and removing the relevant documents from the collection for some queries).